\documentclass[12pt]{article}

\usepackage{graphicx}

\def\beq   {\begin{equation}}
\def\eeq   {\end{equation}}
\def\bea {\begin{eqnarray}}
\def\eea {\end{eqnarray}}

\def\psla{p\kern-.45em/}
\def\ksla{k\kern-.45em/}

\def\sq{\ifmmode{\tilde{q}} \else{$\tilde{q}$} \fi}
\def\sg{\ifmmode{\tilde{g}} \else{$\tilde{g}$} \fi}

\def\dr{\ifmmode{\overline{\rm DR}} \else{$\overline{\rm DR}$} \fi}
\def\ms{\ifmmode{\overline{\rm MS}} \else{$\overline{\rm MS}$} \fi}

\begin{document}


\vspace*{-1cm} 
\begin{flushright}
  TU-848 
\end{flushright}

\vspace*{1.4cm}

\begin{center}

{\Large \bf
Electroweak two-loop contribution to the 
mass splitting within a new heavy SU(2)$_L$ fermion multiplet
}\\

\vspace{10mm}

{\large Youichi Yamada
\footnote{E-mail address: yamada@tuhep.phys.tohoku.ac.jp }}

\vspace{8mm}
\begin{tabular}{l}
{\it Department of Physics, Tohoku University, Sendai 980-8578, Japan}
\end{tabular}

\end{center}

\vspace{3cm}

\begin{abstract}
\baselineskip=15pt
New heavy particles in an SU(2)$_L$ multiplet, sometimes introduced 
in extensions of the standard model, have highly degenerate 
tree-level mass $M$ if their couplings to the 
Higgs bosons are very small or forbidden. 
However, loop corrections may generate 
the gauge-symmetry-breaking mass splitting within the multiplet, 
which does not vanish in the large $M$ limit due to the threshold 
singularity. 
We calculate the electroweak contribution to the mass splitting 
for a heavy fermion multiplet, to the two-loop order. 
Numerically, two-loop electroweak contributions are typically O(MeV). 
\end{abstract}

\vspace{20mm}

\newpage
\pagestyle{plain}
\baselineskip=15pt

\section{Introduction}
In some extensions of the standard model, there are new heavy 
particles which belong to an SU(2)$_L\times$U(1) multiplet $F$ and 
have no, or very small, mixing with other particles. 
The masses of these particles are almost degenerate to 
a value $M$ by the gauge symmetry. 
Although the spontaneous breaking of the SU(2)$_L\times$U(1) 
symmetry by Higgs bosons may generate mass splitting $\delta M$ 
among them, the tree-level mass splitting generally behaves 
as $\delta M\sim m_W^2/M$ and becomes very small for $M\gg m_W$. 
This is especially the case for a very heavy fermion multiplet 
where tree-level renormalizable couplings to the Higgs bosons 
$\bar{F}FH$ are forbidden by symmetry. 
Some of the examples are the almost pure winos or 
higgsinos \cite{mizuta,mizuta2,Chen,Giudice:1995qk,amsb,amsb2,
Gunion:1999jr,hmn}
in special parameter regions of the minimal supersymmetric 
standard model \cite{mssm}, 
SU(2)$_L$ triplet fermions in Type III seesaw model for 
neutrino masses \cite{Foot:1988aq,Ma:2008cu}, and also 
models \cite{extral,mdm} where vector-like heavy fermion multiplets 
are added to the standard model by hand. 

In such cases, it has been known \cite{mizuta2,Giudice:1995qk,amsb2,hmn,
Ma:2008cu,extral,mdm,Cheng:1998hc} that 
the dominant part of the gauge-symmetry-breaking mass 
splitting within the multiplet $F$ 
comes from the radiative correction. 
Although the form of the mass correction strongly depends on 
models, the contributions involving 
electroweak gauge bosons $V=(\gamma, Z, W)$, shown in Fig.~1(a) 
for the one-loop, 
are common in a wide class of extended models. 
Since gauge symmetry breaking in this diagram comes from 
the squared masses ($m_W^2$, $m_Z^2$) in the loops, one naively expect 
the $O(\alpha_2 m_W^2/M)$ contribution to the mass splitting. 
However, due to the singularity of the diagram near the threshold, at 
$p^2=M^2\sim (M+m_V)^2$, $O(\alpha_2 m_W)$ contribution 
to the mass splitting appears, which does not vanish in 
the $M\gg m_W$ limit: Roughly speaking, it is ``nondecoupling''. 
This mass splitting is phenomenologically interesting, 
especially in the case 
where the neutral component $f^0$ of $F$, either fermion or boson, is 
stable or has very long lifetime, and may be a candidate for 
the cosmological dark matter. 
In such a case, the loop-generated mass splitting between 
charged components $f^Q(Q\neq 0)$ of $F$ and $f^0$ is 
crucial for estimating the rates of the $f^Q\to f^0+\cdots$ decays 
expected at colliders, 
and also for possible resonant annihilation $f^0f^0\to f^Qf^{-Q}\to VV$ 
for indirect detection of $f^0$ \cite{hmn,mdm}. 

To evaluate the mass splitting within $F$ to the next-to-leading order, 
we need two-loop calculation of the mass correction for the members 
of $F$. In this paper, we perform such calculation for 
the loop corrections by the standard model particles, 
generated by the electroweak gauge interactions of $F$. 
For simplicity, we concentrate on the SU(2)$_L$-breaking and 
``nondecoupling'' part of the mass correction, which should be 
relevant for the mass splitting in the $M\gg m_W$ case. 

\section{One-loop mass correction}
Since the electroweak contributions to the mass correction 
should be determined by the SU(2)$_L\times$U(1) representation of $F$, 
we work in the framework of the Minimal Dark Matter model \cite{mdm}, 
which has been proposed as a minimalist approach to 
the dark matter problem, for the fermion case. 
In this case, Dirac or Majorana fermions in an SU(2)$_L$ 
multiplet $F$ with SU(2)$_L$ isospin $I$ and U(1) hypercharge $Y$ 
(and having no SU(3) color) are added to the standard model. 
The lagrangian is 
\begin{equation}
{\cal L} = {\cal L}_{\rm SM} + c \bar{F}[\gamma^{\mu} D_{\mu} - M ]F , 
\end{equation}
where $c=1(1/2)$ for Dirac(Majorana) fermions, respectively. 
Note that the mass corrections presented in this paper are common 
to both types of fermions. 
$D_{\mu}$ denotes SU(2)$_L\times$U(1) gauge covariant derivative 
for $F$. Since $F$ has no direct couplings to the Higgs boson, 
the members of $F$, $f^Q$ (with charge $Q=I_3+Y$, $I_3=-I,-I+1,\ldots,I$) 
have a common mass $M$ at the tree-level. 
We assume that $M$ is sufficiently larger than the masses of 
standard model particles ($W$, $Z$, top quark $t$, Higgs boson $h$), 
typically $M=O$(TeV) which is cosmologically favored 
in the Minimal Dark Matter model \cite{mdm}. 
We also use approximation that all other particles in 
the standard model are massless. 

The pole mass $M_p$ of $f^Q$ at the two-loop order is 
given in terms of the self energy of $f^Q$ 
\begin{equation}
\Sigma(p) \equiv \Sigma_K(p^2)\psla + \Sigma_M(p^2) , 
\end{equation}
as 
\begin{eqnarray}
M_p &=& \frac{ M - \Sigma_M(M_p^2) }{ 1 + \Sigma_K(M_p^2) } 
\nonumber\\ 
&=&  M - [ M\Sigma_K^{(1)}(M^2)+\Sigma_M^{(1)}(M^2) ]
- [ M\Sigma_K^{(2)}(M^2)+\Sigma_M^{(2)}(M^2) ]  \nonumber \\ 
&& + [ M\Sigma_K^{(1)}(M^2)+\Sigma_M^{(1)}(M^2) ]
[ \Sigma_K^{(1)}(M^2) + 2M^2\dot{\Sigma}_K^{(1)}(M^2) 
+ 2M\dot{\Sigma}_M^{(1)}(M^2) ] \nonumber \\
& \equiv & M+ \delta M^{(1)}+ \delta M^{(2)} . 
\label{eq:masscor}
\end{eqnarray}
Here $\Sigma_{K,M}^{(1)}$ and $\Sigma_{K,M}^{(2)}$ are 
the one-loop and two-loop parts, respectively. 
The dot in Eq.~(\ref{eq:masscor}) denotes the derivative 
with respect to the external momentum squared. 
The absorptive part of the self energy is $O(g^6)$ and 
need not be considered here. Loop integrals are regularized by the 
dimensional regularization ($D=4-2\epsilon$) with 
the \ms subtraction scheme. 

The form of the one-loop mass correction $\delta M^{(1)}$ 
is well known \cite{mizuta2,amsb2,hmn,Ma:2008cu,extral,mdm,
Cheng:1998hc,pierce}. 
Abbreviating the factor $\alpha_2/(4\pi)$, it is expressed as 
\begin{equation}
\delta M^{(1)} = (C_F-I_3^2)X^{(1)}_W 
+ s_W^2 (I_3+Y)^2 X^{(1)}_{\gamma} 
+ c_W^2 (I_3-t_W^2 Y)^2 X^{(1)}_Z , 
\label{eq:dm1}
\end{equation} 
where $C_F=I(I+1)$, $c_W\equiv\cos\theta_W=m_W/m_Z$, $s_W\equiv\sin\theta_W$, 
$t_W\equiv\tan\theta_W$, and 
\begin{eqnarray}
X_V^{(1)} &=& 
M \left[ \left( 2+\frac{m_V^2}{M^2} \right) B_0(M^2, M, m_V) -1 
+\frac{1}{M^2}\{ A(M)-A(m_V) \} \right] 
\nonumber \\
&=& M \left[ \frac{3}{\epsilon}-3\log M^2 +4 -f\left( \frac{m_V}{M}\right) 
\right] , 
\nonumber \\
f(x) &\equiv & 2x(2+x^2)\sqrt{4-x^2}\tan^{-1}\frac{\sqrt{2-x}}{\sqrt{2+x}}
-x^2+x^4\log x 
\nonumber \\
&=& 2\pi x -3x^2+\frac{3}{4}\pi x^3+ O(x^4) . 
\label{eq:x1}
\end{eqnarray}
We use the Passarino-Veltman one-loop functions \cite{PV} defined as 
\begin{eqnarray}
A(m) &=& 
\frac{1}{\epsilon(1-\epsilon)}(m^2)^{1-\epsilon} \, , 
\nonumber \\
B_0(p^2, m_1, m_2) &=& 
\frac{1}{\epsilon}\int_0^1 dz \, [ (1-z)m_1^2+zm_2^2 - z(1-z)p^2 
-i\delta ]^{-\epsilon},  
\nonumber \\
B_{22}(p^2, m_1, m_2) &=& 
\frac{1}{2\epsilon(1-\epsilon)}\int_0^1 dz \, [ (1-z)m_1^2+zm_2^2 - z(1-z)p^2 
-i\delta]^{1-\epsilon},  
\end{eqnarray}
and 
\begin{equation}
\widetilde{B}_{22}(p^2, m_1, m_2) = B_{22}(p^2, m_1, m_2) 
- \frac{1}{4} [ A(m_1)+A(m_2) ]. 
\end{equation}
The $O(m_V)$ term of Eq.~(\ref{eq:x1}) gives the nondecoupling mass 
splitting within the multiplet. 
For example, for $Y=0$, the one-loop mass splitting between 
$f^Q$ and the neutral component $f^0$ of $F$ is written as \cite{mdm}, 
independent of $I$,  
\begin{equation}
M(f^Q)-M(f^0) = Q^2 \Delta M^{(1)} . 
\label{eq9}
\end{equation}
where, in the $M\gg m_W$ limit, 
\begin{equation}
\Delta M^{(1)} 
=\frac{\alpha_2}{2} (m_W-c_W^2m_Z) = (166.99\pm 0.07) {\rm MeV} .
\label{eq10}
\end{equation}
The numerical value in Eq.~(\ref{eq10}) is obtained by 
using the pole masses $m_W=(80.398\pm 0.025)$ GeV, 
$m_Z=91.1876$ GeV, $\alpha_2=\alpha(m_Z)/s_W^2=\alpha(m_Z)/
(1-m_W^2/m_Z^2)$, 
and the QED running coupling in the \ms scheme 
$\alpha(m_Z)=(127.93\pm 0.03)^{-1}$, 
cited from Ref.~\cite{PDG}, as input parameters. Note that the 
value (\ref{eq10}) should change by $\sim 1$ MeV depending 
on choices of the renormalization scheme for the input parameters. 

\section{Two-loop mass correction}

We now calculate the two-loop mass correction $\delta M^{(2)}$ 
coming from diagrams shown in Fig.~1(b-e). 
We use Feynman gauge fixing for simplicity, although 
the final result should not depend on the gauge fixing method. 

\begin{figure}[ht]
\begin{center}
\includegraphics[width=15cm]{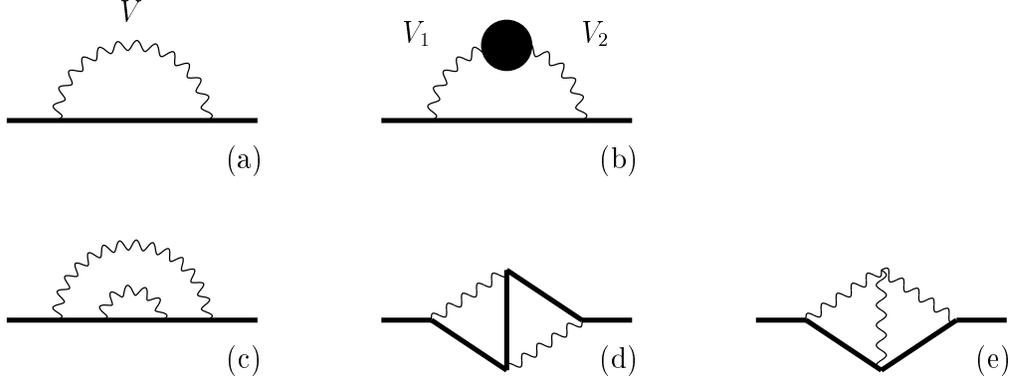}
\end{center}
\caption{ 
\small One-loop (a) and two-loop (b-e) contributions 
to the self energy of the heavy fermions $f$ 
in the multiplet $F$. 
The solid thick line and wavy line represent $F$ 
and electroweak gauge bosons $V=(\gamma,Z,W)$, respectively. 
The black circle in (b) represents the one-loop self energy of 
the gauge bosons $\Pi^{V_1V_2}$, by the standard model particles. 
}
\label{diagram1}
\end{figure}

The contribution of the diagram Fig. 1(b) with the insertion 
of the one-loop self energy of the electroweak gauge boson, 
$\Pi^{V_1V_2}_{\mu\nu}(k)=g_{\mu\nu}\Pi^{V_1V_2}(k^2)+
O(k_{\mu}k_{\nu})$, is written as 
\begin{eqnarray}
\delta M^{(2,1)} &=& -(C_F-I_3^2)\Delta\Sigma_{WW} 
- s_W^2 (I_3+Y)^2\Delta\Sigma_{\gamma\gamma} 
\nonumber \\
&& - 2s_W c_W (I_3+Y)(I_3-t_W^2 Y)\Delta\Sigma_{\gamma Z} 
- c_W^2(I_3-t_W^2 Y)^2\Delta\Sigma_{ZZ} ,
\label{eq11}
\end{eqnarray} 
where 
\begin{eqnarray}
\Delta\Sigma_{V_1V_2} &=& 
\left. 
ig^2 \int\frac{d^Dk}{(2\pi)^D}
\frac{\gamma^{\mu}(\ksla+\psla+M) \gamma_{\mu} 
\Pi^{V_1V_2}(k^2) }
{[k^2-m_{V_1}^2][k^2-m_{V_2}^2] [ (k+p)^2-M^2 ] } 
\right|_{\psla = M} \, .
\label{eq:sigma2b}
\end{eqnarray}
Here we list the analytic forms of $\Pi^{V_1V_2}(k^2)$ in the standard 
model for completeness \cite{SMself}. The contributions from the 
$(t,b)$ quark loops are, up to the overall factor 
$N_c\alpha_2/(4\pi)$ ($N_c=3$ is the color number of quarks), 
\begin{eqnarray}
\Pi_{tb}^{WW}(k^2) 
&=& \frac{1}{2} 
\left[ -4\widetilde{B}_{22}(k_1^2,m_t,0)-(k_1^2-m_t^2)B_0(k_1^2,m_t,0)
\right] , 
\label{eq13}
\\
\Pi_{tb}^{\gamma\gamma}(k^2) &=& \sum_{q=t,b} s_W^2Q_q^2\Pi^{vv}_q(k^2) , \\
\Pi_{tb}^{\gamma Z}(k^2) &=& \sum_{q=t,b} 
\frac{s_W}{c_W} Q_q\left( \frac{1}{2}I_{3q}-Q_q s_W^2
\right) \Pi^{vv}_q(k^2) , \\
\Pi_{tb}^{ZZ}(k^2) &=& 
\sum_{q=t,b} \frac{1}{c_W^2} 
\left( \frac{1}{2}(I_{3q})^2  -I_{3q}s_W^2Q_q 
+s_W^4 Q_q^2 \right) \Pi^{vv}_q(k^2) 
\nonumber\\ 
&& +\frac{m_t^2 }{2c_W^2} B_0(k^2,m_t,m_t) , 
\end{eqnarray}
where 
\begin{equation}
\Pi^{vv}_{q}(k^2) \equiv 
-8\widetilde{B}_{22}(k_1^2,m_q,m_q)-2k_1^2B_0(k_1^2,m_q,m_q) .
\label{eq17}
\end{equation}
The contributions of other quarks and leptons are obtained by appropriate 
changes of $m_t$, $Q_q$, and $N_c$. 
For the gauge and Higgs boson loops, we have, abbreviating 
the overall factor $\alpha_2/(4\pi)$, 
\begin{eqnarray}
\Pi_{Vh}^{WW}(k^2) &=& 
s_W^2[ 8(1-\epsilon)\widetilde{B}_{22}(k^2,m_W,0)+4k^2 B_0(k^2,m_W,0) ] 
\nonumber\\ 
&& +(1+8(1-\epsilon)c_W^2) \widetilde{B}_{22}(k^2,m_W,m_Z) 
\nonumber\\ 
&& +[4c_W^2 k^2 -m_Z^2 + 3m_W^2]B_0(k^2,m_W,m_Z) 
\nonumber\\ 
&& +\widetilde{B}_{22}(k^2,m_W,m_h) -m_W^2 B_0(k^2,m_W,m_h) , 
\label{eq18}
\\
\Pi_{Vh}^{\gamma\gamma}(k^2) &=& 
s_W^2 [ 4(3-2\epsilon)\widetilde{B}_{22}(k^2,m_W,m_W) 
+4k^2 B_0(k^2,m_W,m_W) ] , 
\\
\Pi_{Vh}^{\gamma Z}(k^2) &=& 
4s_Wc_W \left( 3-2\epsilon-\frac{1}{2c_W^2} \right) 
\widetilde{B}_{22}(k^2,m_W,m_W) 
\nonumber\\
&& +s_Wc_W ( 4k^2+2m_Z^2 ) B_0(k^2,m_W,m_W) , 
\\
\Pi_{Vh}^{ZZ}(k^2) &=& 
\left[ 4c_W^2(3-2\epsilon)-4+\frac{1}{c_W^2} \right]
\widetilde{B}_{22}(k^2,m_W,m_W) 
\nonumber\\
&& +[ 4c_W^2 k^2+ 4m_W^2 - 2 m_Z^2  ]
B_0(k^2,m_W,m_W)
\nonumber \\
&& +\frac{1}{c_W^2} [ 
\widetilde{B}_{22}(k^2,m_Z,m_h) -m_Z^2 B_0(k^2,m_Z,m_h) ] .
\label{eq21}
\end{eqnarray}
In addition, there are also the contributions of $F$ 
to $\Pi^{V_1V_2}$. However, it is shown that 
the resulting $O(m_W)$ contributions to $\delta M^{(2)}$ are completely 
cancelled by the renormalization of the parameters in 
$\delta M^{(1)}$. 

We may calculate the integrals (\ref{eq:sigma2b}) 
by extending the general formulas for the two-loop mass 
corrections \cite{martin}, by including finite masses for 
$(W,Z)$. However, since we are interested in the SU(2)$_L$-breaking 
and nondecoupling part of Eq.~(\ref{eq:sigma2b}), it is 
prefered to expand the integrals (\ref{eq:sigma2b}) 
in $m_W(\sim m_Z,m_t,m_h)$ and then separate the $O(m_W)$ terms 
from the dominant and gauge-symmetric $O(M)$ terms, 
before numerical evaluation. 
This is achieved by applying the asymptotic expansion of the 
Feynman integrals near the threshold $p^2=M^2$, 
as described in Ref.~\cite{THE}. The $O(m_W)$ part of the 
integral (\ref{eq:sigma2b}) is then obtained as 
\begin{eqnarray}
\Delta\Sigma_{V_1V_2} |_{O(m_W)} 
&\to &
ig^2 \int\frac{d^Dk}{(2\pi)^D}
\frac{2M}{[k^2-m_{V_1}^2][k^2-m_{V_2}^2] (2k\cdot p) }
\Pi^{V_1V_2}(k^2) . 
\label{eq:sigma2b2} 
\end{eqnarray}
In the following, we show only the $O(m_W)$ part (\ref{eq:sigma2b2}) 
of the corrections $\Delta\Sigma_{V_1V_2}$.  
By substituting the self energies (\ref{eq13}--\ref{eq21}), 
the integrals (\ref{eq:sigma2b2}) 
are expressed in terms of the two-loop functions ($a=1,2$) 
\begin{eqnarray}
\frac{i}{(4\pi)^2} X_{0-a}(m_V, m_1, m_2) &\equiv & 
\int\frac{d^Dk}{(2\pi)^D}
\frac{M}{[k^2-m_V^2]^a (2k\cdot p) }
\left[ B_0(k^2, m_1, m_2) -\frac{1}{\epsilon} \right] , 
\nonumber \\
\frac{i}{(4\pi)^2} X_{22-a}(m_V, m_1, m_2) &\equiv & 
\int\frac{d^Dk}{(2\pi)^D}
\frac{M}{[k^2-m_V^2]^a (2k\cdot p) }
\nonumber \\ 
&& \times \left[ B_{22}(k^2, m_1, m_2) -\frac{1}{\epsilon}
\left( \frac{m_1^2+m_2^2}{4}- \frac{k^2}{12} \right) \right] , 
\label{xinteg}
\end{eqnarray}
and products of the one-loop functions. Note that the functions in 
Eq.~(\ref{xinteg}) are independent of $M$ and have no overall 
divergences. We calculate these 
functions by numerical integration of the Feynman parameter integrals 
shown below, 
\begin{eqnarray}
X_{0-1}(m_V, m_1, m_2) &=& \pi m_V 
\left[ 
\log\frac{m_V^2}{\mu^2} 
\right. 
\nonumber \\ && 
\left. 
- \int_0^1 dz \, 
\left( 2 \frac{\sqrt{r_1(1-z)+r_2z}}{\sqrt{z(1-z)}} 
-2\log(\sqrt{r_1(1-z)+r_2z}+\sqrt{z(1-z)} ) 
\right) 
\right] ,
\nonumber \\
\\
X_{0-2}(m_V, m_1, m_2) &=& \frac{\pi}{2m_V}
\left[ \log\frac{m_V^2}{\mu^2}
+2\int_0^1 dz \, \left( 
\frac{\sqrt{z(1-z)}}{\sqrt{r_1(1-z)+r_2z}+\sqrt{z(1-z)} } 
\right. \right. 
\nonumber \\ && 
\left. \left. +\log(\sqrt{r_1(1-z)+r_2z}+\sqrt{z(1-z)} ) 
\right) 
\right] , 
\\
X_{22-1}(m_V, m_1, m_2) &=& -\frac{\pi}{3}m_V^3 
\left[  \frac{1}{4}\{ 1-3(r_1+r_2) \} \log\frac{m_V^2}{\mu^2}
-\frac{2}{3} +\frac{9}{4}(r_1+r_2) 
\right. 
\nonumber \\ && 
\left. 
+\int_0^1 dz \, \left( 
\{ -3z(1-z)+2(1-z)r_1+2zr_2 \}\frac{\sqrt{(1-z)r_1+zr_2} }{\sqrt{z(1-z)} } 
\right. \right.
\nonumber\\ && 
+3\{ z(1-z)-(1-z)r_1-zr_2 \} \left\{ 
\log(\sqrt{r_1(1-z)+r_2z}+\sqrt{z(1-z)} )
\right. 
\nonumber \\ && 
\left. \left. \left. 
-\frac{1}{2}\log{z(1-z)} 
\right\}
\right) \right] ,
\\
X_{22-2}(m_V, m_1, m_2) &=& -\frac{\pi}{2}m_V 
\left[ 
\frac{1}{4}(1-r_1-r_2)\log\frac{m_V^2}{\mu^2} 
-\frac{1}{2} +\frac{3}{4}(r_1+r_2) 
\right. 
\nonumber \\ && 
+\int_0^1 dz \, \left( 
-3\sqrt{z(1-z)}\sqrt{(1-z)r_1+zr_2} 
+\{ 3z(1-z)-(1-z)r_1-zr_2 \}
\right. 
\nonumber \\ && 
\left. \left. 
\times \left\{ 
\log(\sqrt{(1-z)r_1+zr_2}+\sqrt{z(1-z)} )- \frac{1}{2}\log(z(1-z))
\right\} \right) \right] ,
\end{eqnarray}
where $r_{1,2}\equiv m_{1,2}^2/m_V^2$. $\mu$ is the 
\ms renormalization scale. 

Here we show the explicit forms of the integrals (\ref{eq:sigma2b2}), 
after subtracting $O(1/\epsilon)$ divergences 
from $\Pi_{V_1V_2}$ by the \ms scheme, and 
separating the mass corrections to the gauge bosons 
$\delta m_V^2=-{\rm Re}\Pi^{VV}(m_V^2)(V=W,Z)$ from $\Pi^{VV}(k^2)$. 
The $(t,b)$ contributions coming from Eqs.~(\ref{eq13}--\ref{eq17})
are, up to the overall factor $N_c\alpha_2/(4\pi)$, 
\begin{eqnarray}
\Delta\Sigma_{WW}^{tb} & = & \frac{1}{2} 
X_{WW}(m_t) -\frac{\pi}{m_W}\delta m_W^{2(tb)}, 
\label{eq27}
\\
s_W^2\Delta\Sigma_{\gamma\gamma}^{tb} & = & 
-4 s_W^4 Q_t^2 \pi^2 m_t  , 
\\
2s_Wc_W\Delta\Sigma_{\gamma Z}^{tb} 
& = & 
2 s_W^2 Q_t(I_{3t}-2Q_t s_W^2) X_{\gamma Z}(m_t) 
\nonumber\\ && 
+2 s_W^2 Q_b(I_{3b}-2Q_b s_W^2) X_{\gamma Z}(0) ,
\\
c_W^2\Delta\Sigma_{ZZ}^{tb} 
& = & 
[ (I_{3t})^2-2I_{3t}Q_ts_W^2+2Q_t^2s_W^4 ] X_{ZZ}(m_t)
-m_t^2 G_0(m_Z,m_t,m_t) 
\nonumber \\ && 
+ [ (I_{3b})^2-2I_{3b}Q_bs_W^2+2Q_b^2s_W^4 ] X_{ZZ}(0)
-c_W^2\frac{\pi}{m_Z}\delta m_Z^{2(tb)}, 
\label{eq30}
\end{eqnarray}
with 
\begin{eqnarray}
X_{WW}(m_t) &=& 8G_{22}(m_W,m_t,0)+2X_{0-1}(m_W,m_t,0)
\nonumber \\ && 
+2(m_W^2-m_t^2)G_0(m_W,m_t,0), 
\\
X_{\gamma Z}(m_t) &=& 
\frac{8}{m_Z^2}X_{22-1}(m_Z,m_q,m_q)+\frac{16\pi^2}{3}\frac{m_t^3}{m_Z^2} 
\nonumber\\ && 
+2X_{0-1}(m_Z,m_t,m_t) + \frac{4\pi m_t^2}{m_Z} 
\left( 1-\log\frac{m_t^2}{\mu^2} \right) , 
\\ 
X_{\gamma Z}(0) &=& \pi m_Z \left(
\frac{4}{3}\log\frac{m_Z^2}{\mu^2} -\frac{20}{9} \right) , 
\\ 
X_{ZZ}(m_t) &=& 
8G_{22}(m_Z,m_t,m_t)
\nonumber \\ && 
+2X_{0-1}(m_Z,m_t,m_t) +2m_Z^2 G_0(m_Z,m_t,m_t) 
\\
X_{ZZ}(0) &=& 
\pi m_Z \left(\frac{4}{3}\log\frac{m_Z^2}{\mu^2} -\frac{8}{9} \right) . 
\end{eqnarray}
Here we used the notations 
\begin{eqnarray}
G_{22}(m_V,m_1,m_2) &\equiv & X_{22-2}(m_V,m_1,m_2)
\nonumber \\ && 
+\frac{\pi}{2m_V}[ {\rm Re}B_{22}(m_V,m_1,m_2) 
-((m_1^2+m_2^2)/4-m_V^2/12)/\epsilon ], 
\nonumber \\ 
G_0 (m_V,m_1,m_2) &\equiv & X_{0-2}(m_V,m_1,m_2)
+\frac{\pi}{2m_V}[ {\rm Re}B_0(m_V,m_1,m_2) - 1/\epsilon ],  
\end{eqnarray}
and substituted analytic forms of the two-loop integrals 
(\ref{xinteg}) at $m_1=m_2=0$ and 
at $m_V\to 0$. Analytic forms of other integrals involving 
$m_t$ are shown in Appendix. 
Contributions of other quarks and leptons are obtained 
by taking $m_t\to 0$, where 
\begin{equation}
X_{WW}(0)= \pi m_W 
\left(\frac{4}{3}\log\frac{m_W^2}{\mu^2}-\frac{8}{9} \right), 
\end{equation}
and, for leptons, changing ($Q_q$, $N_c$). 
Similarly, the gauge and Higgs boson 
contributions coming from Eqs.~(\ref{eq18}--\ref{eq21}) are, 
up to the factor $\alpha_2/(4\pi)$, 
\begin{eqnarray}
\Delta\Sigma_{WW}^{Vh} &=& 
-16s_W^2G_{22}(m_W,m_W,0)
-8s_W^2( X_{0-1}(m_W,m_W,0)+m_W^2 G_0(m_W^2,m_W,0) ) 
\nonumber\\ && 
-2(8c_W^2+1) G_{22}(m_W,m_W,m_Z)
+\frac{4}{3}\pi m_W 
-8 c_W^2 X_{0-1}(m_W,m_W,m_Z)
\nonumber\\ && 
-2((5m_W^2+m_Z^2)c_W^2-m_Z^2s_W^4) G_0(m_W,m_W,m_Z) 
\nonumber \\
&& -2G_{22}(m_W,m_W,m_h) +2m_W^2 G_0(m_W,m_W,m_h) 
-\frac{\pi}{m_W}\delta m_W^{2(Vh)} , 
\label{eq38}
\\
s_W^2\Delta\Sigma_{\gamma\gamma}^{Vh} 
&=& 10\pi^2 s_W^4 m_W, 
\\
2s_Wc_W\Delta\Sigma_{\gamma Z}^{Vh} &=& 
\frac{8s_W^2(6c_W^2-1)}{m_Z^2}
[ -X_{22-1}(m_Z,m_W,m_W)  -\frac{2}{3}\pi^2m_W^3 
-\frac{\pi}{2} m_Z m_W^2(1-\log \frac{m_W^2}{\mu^2}) ] 
\nonumber \\ && 
+\frac{8}{3}s_W^2c_W^2 \pi m_Z 
-8s_W^2c_W^2[3X_{0-1}(m_Z,m_W,m_W)+2\pi^2 m_W], 
\\
c_W^2\Delta\Sigma_{ZZ}^{Vh} &=& 
-2(12c_W^4-4c_W^2+1) G_{22}(m_Z,m_W,m_W)
+\frac{4}{3} c_W^4 \pi m_Z  
\nonumber\\
&& -8c_W^4 X_{0-1}(m_Z,m_W,m_W) 
-4(4c_W^2-1)m_W^2 [ G_0(m_Z,m_W,m_W)]
\nonumber\\ 
&& -2G_{22}(m_Z,m_Z,m_h) +2m_Z^2 G_0(m_Z,m_Z,m_h) 
-c_W^2\frac{\pi}{m_Z}\delta m_Z^{2(Vh)} . 
\label{eq41}
\end{eqnarray}


Other diagrams shown in Fig. 1(c--e) are also evaluated by using 
the threshold expansion \cite{THE}, keeping only the $O(m_V)$ parts. 
Their sum, with subtracting subdivergences by the \ms scheme and 
after the (one-loop)$\times$(one-loop) term 
in Eq. (\ref{eq:masscor}) is added, is given as 
\begin{eqnarray}
\delta M^{(2,2)} &=& 4\pi m_W (C_F-I_3^2)
\left[ c_W^2 \log \frac{m_Z^2}{\mu^2} + (2-c_W^2)\log\frac{m_W^2}{\mu^2} 
+ 4 s_W^2 (-1+\log 2) \right]
\nonumber\\ 
&& + 8\pi c_W^2 m_Z I_3(I_3 - t_W^2 Y)\log \frac{m_W^2}{\mu^2} 
- 4\pi c_W^2 (C_F-I_3^2) f_{ZW} . 
\label{eq42}
\end{eqnarray}
Here 
\begin{eqnarray}
f_{ZW} &\equiv & 
-\frac{1}{3}( 2+c_W^2)m_W \int_0^1 dz \, z^{-3/2} (1-z)^{-1/2}
[ (c_W^{-2} z +1-z)^{3/2}-1 ] 
\nonumber\\ 
&&  -\frac{1}{3}( 2+c_W^{-2})m_Z \int_0^1 dz \, z^{-3/2} (1-z)^{-1/2}
[ (c_W^2 z +1-z)^{3/2}-1 ] 
\nonumber \\
&\sim & -0.027 m_W , 
\end{eqnarray}
is the two-loop function appearing in Fig.~1(c,d) with both $W$ and $Z$ 
bosons. 

We then need to add the counterterms coming from the 
renormalization of the parameters in 
the one-loop contributions (\ref{eq:dm1}, \ref{eq:x1}); 
$(m_W,m_Z)$ in $X^{(1)}_{W,Z}$ and ($\alpha_2$, $c_W^2$, \ldots) 
in the coupling constants. 
We adopt the scheme where the pole masses ($m_W$, $m_Z$) and 
the \ms running coupling of QED $\alpha(m_Z)$, which are used 
in Eq.~(\ref{eq10}), are chosen as the input parameters. 
In this scheme, the renormalization is achieved by 
removing the last $O(\delta m_V^2)$ terms from $\Delta\Sigma_{WW}$ 
(\ref{eq27}, \ref{eq38}) and $\Delta\Sigma_{ZZ}$ (\ref{eq30}, \ref{eq41}), 
and adding the counterterms for ($\alpha_2$, $c_W^2$, \ldots) expressed as 
tree-level functions of ($m_Z$, $m_W$, $\alpha(m_Z)$). 
It is checked that the final form of the two-loop $O(m_W)$ 
mass correction to $f^Q$ is finite and 
independent of the \ms renormalization scale $\mu$. 

Here we comment on the mass splitting of a new heavy 
scalar SU(2)$_L$ multiplet $S$. In contrast to 
the case of the fermion multiplet, 
direct couplings of $S$ to the Higgs bosons, 
such as $S^*SH^{\dagger}H$, should always 
exist \cite{mdm,Hambye:2009pw}. 
Nevertheless, assuming that the effect of these 
direct couplings is negligible, 
we have verified that the nondecoupling $O(m_W)$ parts of the 
one-loop \cite{mdm,Yamada:1996jf} and two-loop mass 
corrections $\delta M$ are identical to those for the fermions 
in the same gauge representation. This result is quite natural 
in the view that the $O(m_W)$ mass correction could be understood 
as the energy of the electroweak gauge fields around a static 
point source, and should be insensitive to the spin of the source 
particle \cite{mdm}. 

\section{Numerical results}
We show the numerical results of the two-loop contributions 
to the mass splitting within the $Y=0$ fermion multiplet. As seen in 
Eqs.~(\ref{eq11}, \ref{eq42}), the one-loop relation (\ref{eq9}) 
still holds with the change 
$\Delta M^{(1)}\to\Delta M^{(1)}+\Delta M^{(2)}$, where 
$\Delta M^{(2)}=\Delta M^{(2,ql)}+\Delta M^{(2,Vh)}$. 

The contribution $\Delta M^{(2,ql)}$ of the quark-lepton subloop 
diagrams (including corresponding counterterms) 
is shown in Fig.~2 as a function of 
$m_t$. At $m_t=171$ GeV, there is cancellation 
between the $(t,b)$ subloop contribution, shown in the dashed line, 
and remaining contribution with subloops of other quarks or leptons, 
$-3.3$ MeV, giving the total shift $-1.5$ MeV at $m_t=171$ GeV. 
\begin{figure}[ht]
\begin{center}
\includegraphics[width=14cm]{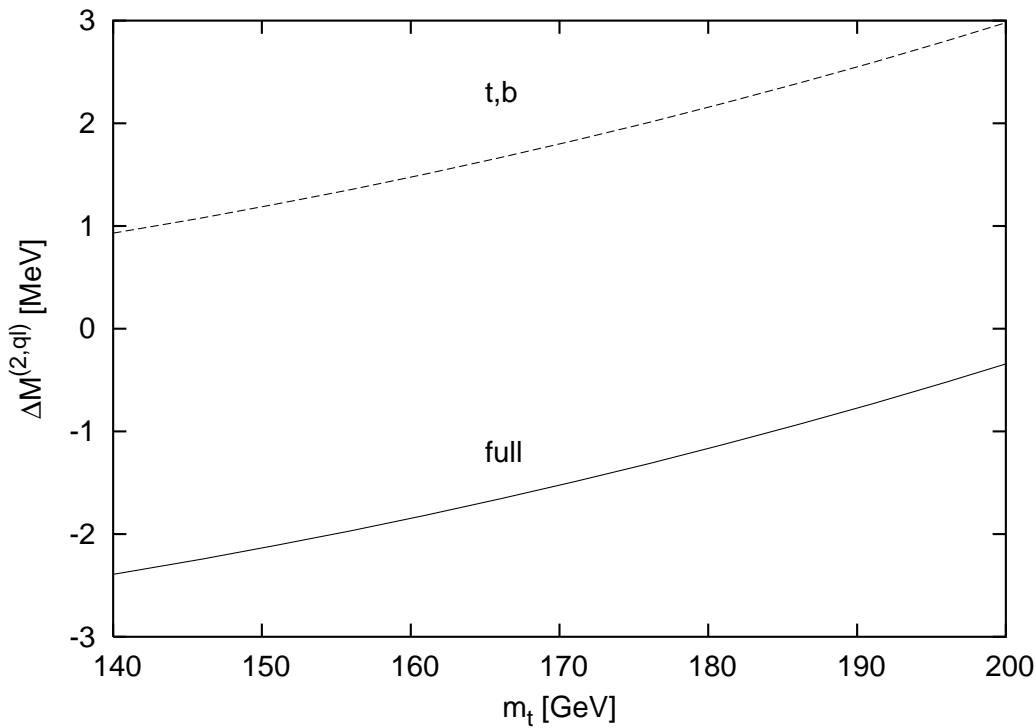}
\end{center}
\caption{ 
\small Two-loop contribution to the mass splitting 
$\Delta M^{(2,ql)}$ between fermions in a heavy 
SU(2)$_L$ multiplet with $Y=0$, from diagrams in Fig.~1(b) 
with quark and lepton subloops. 
Solid and dashed lines denote 
full and $(t,b)$ subloop contributions, respectively. 
}
\label{fig1}
\end{figure}

The remaining contribution $\Delta M^{(2,Vh)}$ from diagrams without 
quarks or leptons (again including corresponding counterterms)
is shown in Fig.~3 as a function of $m_h$. At $m_h=140$~GeV, the shift is 
$-0.9$ MeV, smaller than the quark-lepton loops. 
\begin{figure}[ht]
\begin{center}
\includegraphics[width=14cm]{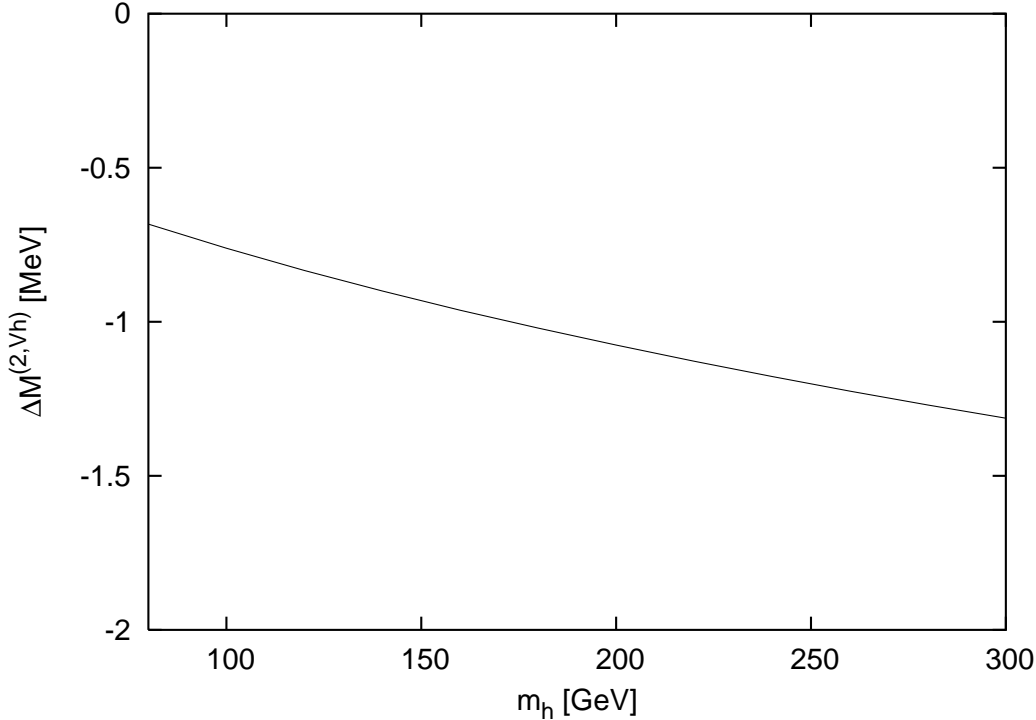}
\end{center}
\caption{ 
\small Two-loop contribution to the mass splitting 
$\Delta M^{(2,Vh)}$ between fermions in a heavy 
SU(2)$_L$ multiplet with $Y=0$, from diagrams in Fig.~1(b-e) 
with gauge and Higgs bosons. 
}
\label{fig2}
\end{figure}

These two-loop contributions are much smaller than the $O(m_W)$ 
part of the leading one-loop contribution (\ref{eq:dm1}), as expected. 
However, for $Y=0$, it may compete with the $M$-dependence of 
the one-loop contribution (\ref{eq9}) 
which behaves like $-0.5(1\,{\rm TeV}/M)^2$ MeV for large $M$ 
due to the accidental cancellation of the $O(m_W^2/M)$ term in 
Eq.~(\ref{eq9}). 
In comparison, in the case of the higgsino-like doublet $F=(f^+,f^0)$ 
with ($I=1/2$, $Y=1/2$), the two-loop corrections to the mass 
splitting $M(f^+)-M(f^0)$, which is $\alpha m_Z/2=356.4$~MeV 
at the one-loop, is $-1.2$~MeV from 
quark and lepton loops at $m_t=171$ GeV, and $-1.8$~MeV from 
gauge and Higgs boson loops at $m_h=140$ GeV, respectively. 

\section{Conclusion}
We have calculated the two-loop electroweak 
contribution to the $O(m_W)$ correction to the masses of 
new heavy fermions in an SU(2)$_L$ multiplet $F$, which causes 
gauge-symmetry-breaking and ``nondecoupling'' mass splitting within $F$. 
Analytic formula of the $O(m_W)$ mass corrections have been presented
for $F$ in general SU(2)$_L\times$U(1) representation. 
The two-loop contribution has turned out to be typically O(MeV), 
which is of similar order to the $M$ dependence 
of the one-loop contribution for the $Y=0$ case. 

\section*{Appendix}
\setcounter{equation}{0}
\renewcommand{\theequation}{A.\arabic{equation}}
In the case of the $(t,b)$ contributions (\ref{eq27}--\ref{eq30}), 
Feynman parameter integrals for the functions (\ref{xinteg}) 
can be analytically performed. 
For $m_1=m_2\equiv \sqrt{r}m_V$ with $r>1/4$, we have 
\begin{eqnarray}
X_{0-1} &=& 
\pi m_V 
\left[ 
\log\frac{m_V^2}{\mu^2} 
- 2+\log r -2\sqrt{4r-1}\tan^{-1}\sqrt{4r-1} \right] ,
\\
X_{0-2} &=& 
\frac{\pi}{2m_V}
\left[ \log\frac{m_V^2}{\mu^2}
+\log r +\frac{2}{\sqrt{4r-1}}\tan^{-1}\sqrt{4r-1} 
\right] , 
\\
X_{22-1} &=& 
-\frac{\pi}{3}m_V^3 
\left[  \frac{1}{4}( 1-6r ) \log\frac{m_V^2}{\mu^2}
-\frac{2}{3} +\frac{7}{2}r 
\right. 
\nonumber \\ && 
\left.
+\frac{1-6r}{4}\log r 
+\frac{1}{2}(4r-1)^{3/2} \tan^{-1}\sqrt{4r-1}
\right] ,
\\
X_{22-2} &=& 
-\frac{\pi}{2}m_V 
\left[ 
\frac{1}{4}(1-2r)\log\frac{m_V^2}{\mu^2} 
-\frac{1}{2} +\frac{1}{2}r 
\right. 
\nonumber \\ && 
\left. 
+\frac{1-2r}{4}\log r 
-\frac{1}{2}\sqrt{4r-1}\tan^{-1}\sqrt{4r-1} 
\right] .
\end{eqnarray}
For $m_2=0$ and $m_1\equiv \sqrt{r}m_V$, 
\begin{eqnarray}
X_{0-1} &=& 
\pi m_V 
\left[ 
\log\frac{m_V^2}{\mu^2} 
-2 -2\sqrt{r}+r\log r - 2(r-1)\log(1+\sqrt{r}) 
\right] ,
\\
X_{0-2} &=& 
\frac{\pi}{2m_V}
\left[ \log\frac{m_V^2}{\mu^2}
-2\sqrt{r} -r\log r +2(r+1)\log(1+\sqrt{r})
\right] , 
\\
X_{22-1} &=& 
-\frac{\pi}{3}m_V^3 
\left[  \frac{1}{4}( 1-3r) \log\frac{m_V^2}{\mu^2}
+\frac{1}{12}(-8-6\sqrt{r} +21r
+16r^{3/2} -3r^2 +6r^{5/2} )
\right. 
\nonumber \\ && 
\left. 
+\frac{(r-3)r^2}{4}\log r -\frac{1}{2}(r-1)^3\log(1+\sqrt{r}) 
\right] ,
\\
X_{22-2} &=& -\frac{\pi}{2}m_V 
\left[ 
\frac{1}{4}(1-r)\log\frac{m_V^2}{\mu^2} 
+\frac{1}{12}(-6 -6\sqrt{r} +3r 
+4r^{3/2} +3r^2 -6r^{5/2} )
\right. 
\nonumber \\ && 
\left. 
-\frac{r^2(r-1)}{4}\log r +\frac{1}{2}(r-1)(r^2-1)\log(1+\sqrt{r}) 
\right] .
\end{eqnarray}


\end{document}